\documentstyle[12pt]{article}
\begin{document}

\begin{center}
{\Large\bf Phase space structure
 and the path integral\\

\vskip 0.2cm
for gauge theories on a cylinder}

\vskip 0.5cm
{\large\bf Sergey V. SHABANOV}$^{ ^{ ^*}}$

\vskip 0.5cm
{\em Service de Physique Theorique de Saclay\\
91191 Gif-sur-Yvette Cedex, France}
\end{center}

\begin{abstract}
The physical phase space  of gauge field theories on a
cylindrical spacetime with an arbitrary compact simple gauge group
is shown to be the quotient ${\bf R}^{2r}/W_A,\ r$ a rank of the gauge
group, $W_A$ the affine Weyl group.
The PI formula resulting from Dirac's operator method contains
a symmetrization with respect to $W_A$
rather than the integration domain reduction. It gives a natural
solution to Gribov's problem. Some features of fermion
quantum dynamics caused by the nontrivial phase space geometry
are briefly discussed.
\end{abstract}

{\bf 1}. A main feature of gauge theories is the  existence of unphysical
variables whose evolution is determined by arbitrary functions of time
\cite{dir}, while physical quantities appear to be independent of the
gauge arbitrariness. Dynamics of physical variables occurs in a
configuration (or phase) space and, therefore,
geometry of the physical configuration (or phase) space
(below denoted as $CS_{ph}$ or $PS_{ph}$, respectively) plays an
important role in the dynamical description. For example, compare particle
dynamics on a circle or in a line, or on $PS_{ph}$ being a sphere. The
classical as well as quantum theories are obviously different.

In the present letter, we analyze $PS_{ph}$ for 2D topological gauge
theories \cite{wit} on a cylindrical spacetime, ${\bf R}\otimes S^1$,
\cite{raj}.  We include also fermion fields in the theory
\cite{sem},\cite{mic}
and observe that their ${PS_{ph}}$ is also modified, which leads to some
dynamical consequences. A main purpose of our work is to construct PI
over $PS_{ph}$ (being different from an Euclidean space) which results
from the Dirac operator approach.

$PS_{ph}$ can be determined as the quotient of the constraint surface in
the whole PS by gauge transformations $G$ generated by all
first-class constraints $\sigma _a$
\begin{equation}
PS_{ph}=PS\vert_{\sigma _a=0}/G\ .
\end{equation}
As has been pointed out in \cite{lv}, $PS_{ph}$ in gauge models may
differ from an ordinary Euclidean space and be, for example, a cone
unfoldable into a half-plane (for a review see \cite{uf}). The path
integral representation of quantum theory depends on the $PS_{ph}$
geometry \cite{uf}-\cite{weyl}, which leads to physical consequences
for gauge field dynamics \cite{uf}, \cite{gf} and a minisuperspace
cosmology \cite{cos}.

\noindent
$\overline{\hspace*{8cm}}$

$\ ^*${\small
On leave of absence from: {\em Laboratory of Theoretical Physics,
J.I.N.R., P.O. Box 79, Moscow, Russia}.}
\newpage
Though the definition (1) is independent of a coordinate
(parametrization) choice and
explicitly gauge-invariant, we need to introduce coordinates on $PS_{ph}$
upon the PI construction (an attempt to define a "coordinate-free" PI has
been proposed in \cite{kla} for non-constrained systems). The
parametrization
choice is motivated by physical reasons. For instance,
in gauge theories one may describe physical degrees of freedom by
transverse potentials ${\bf A}^\perp$ and their conjugated momenta
${\bf E}^\perp$. In QED, there is a one-to-one correspondence between
$PS_{ph}$ and $[{\bf A}^\perp]\otimes [{\bf E}^\perp]\equiv PS^\perp\
([{\bf A}^\perp]$ implies the functional space of all configurations
${\bf A}^\perp$), i.e. $PS_{ph}\sim PS^\perp$. However, for non-Abelian
theories $PS_{ph}$ does not coincide with $PS^\perp$ because there are
gauge equivalent configurations in $[{\bf A}^\perp]$, Gribov's copies
\cite{gr}. Moreover, this parametrization (or gauge fixing) ambiguity
always arises and has a geometric nature \cite{sin} related to topological
properties of $PS_{ph}$. Notice that Gribov's copies themselves do not
have much physical meaning because they are strongly connected with a
concrete choice of gauge fixing condition (or parametrizing $PS_{ph}$)
which is rather arbitrary, while a topology of $PS_{ph}$ is
gauge-independent.

2D topological gauge theories in the Hamiltonian approach (meaning a
cylindrical spacetime) have a finite number of physical degrees of
freedom and may serve as good toy models for verifying some ideas and
methods invented for 4D gauge theories. Recently, we have proposed a
PI construction method for any reasonable parametrization of $PS_{ph}$
(for any gauge fixing condition) which is based on the Dirac formalism
of quantizing first-class constrained systems
 \cite{gf}, \cite{lec}. Bellow we shall
describe its main points for the Yang-Mills theory and then apply it to
2D gauge theories. The path integral appears to be modified as compared
with the path integral formalism constructed by means of the Faddeev-Popov
trick, but it recovers all results obtained by the loop (gauge-fixing free)
 approach \cite{raj},\cite{mig}.

{\bf 2}. Let us turn directly to establishing $PS_{ph}$ in the Yang-Mills
theory on a cylindrical spacetime. The Hamiltonian and constraint read
\begin{eqnarray}
H&=&\frac 12\int\limits_{0}^{2\pi l}dx(E,E)\ ,\\
\sigma &=&\nabla (A)E=\partial E +g[A,E]=0\ ,
\end{eqnarray}
respectively. Here $l$ is a radius of space, $E$ and $A$ are the colour
electric field and potential, they are elements of a Lie algebra $X$ of a
simple compact group $G$; components of $E$ and $A$ serve as canonically
conjugated variables. The brackets $(,)$  and $[,]$ stand for an
invariant inner product and a commutator in $X$, respectively; $\partial
\equiv\partial /\partial x$, and $g$ is a coupling constant. The
constraint (3) generates gauge transformations
\begin{equation}
E\rightarrow \Omega E\Omega^{-1}\ ,\ \ \ \ \ \
A\rightarrow \Omega A \Omega^{-1}+
g^{-1}\Omega\partial\Omega^{-1}=A^\Omega
\end{equation}
with $\Omega=\Omega(x)$ being an element of the gauge group $G$.

As all field variables are functions on $S^1$, they should be periodic
with a period $2\pi l$. Also, $\Omega(x+2\pi l)=\Omega(x)$ (modulo the
group center). We denote
the space of functions on $S^1$ as ${\cal F}[S^1]$; any element $f\in
{\cal F}[S^1]$ can be decomposed into the Fourier series
\begin{equation}
f(x)=f_0 +\sum\limits_{n=1}^{\infty}\left(f_{s,n}\sin\frac{nx}{l}
+f_{c,n}\cos\frac{nx}{l}\right)\ .
\end{equation}
Formally, any configuration $A$ might be reduced to zero by the
transformation (4) with $\Omega^{-1}=\Omega _W(x)=P\exp(-g\int_0^xdyA(y))$
since $(\partial +gA)\Omega_W=0$. But the group element $\Omega _W$ does
not belong to the gauge group because
\begin{equation}
\Omega _W(x+2\pi l)=W[A]\Omega _W(x)\ ,
\end{equation}
where $W[A]=\Omega _W(2\pi l)$ is the Wilson loop. Put, for example,
$A=A_0=const$ then, $W[A_0]=\exp(-2\pi glA_0)\neq 1$. Therefore, the
system possesses physical degrees of freedom.
To separate them, we first reduce fields $E(x)$ and $A(x)$ on the
constraint surface (3) to constant
configurations $E_0$ and $A_0$ by means of a gauge transformation
\cite{raj}. The residual continuous gauge arbitrariness consists of
constant gauge transformations of $E_0,\ A_0\in X$. The surface (3) is
reduced to $[E_0,A_0]=0$.

Any element of $X$ can be represented in the form \cite{hel} $A_0=\Omega
_A h\Omega _A^{-1},\ h$ an element of the Cartan subalgebra $H$ in $X$,
$\Omega _A\in G$. Therefore, any phase-space point $E_0,A_0$ on the
surface $[E_0,A_0]=0$ can be obtained from the pair $p_h,h\in H$ by a gauge
transformation $\Omega =\Omega _A$. Indeed, $[\Omega _A^{-1}E_0\Omega _A,h]
\equiv [p_h,h]=0$, i.e., $p_h\in H$. The element $h$ has a stationary group
being the Cartan subgroup $G_H$ in $G$. This means that not all
of the constraints (3) are independent (Eq.(3) implies an infinite number of
constraints since $\sigma \in {\cal F}[S^1]$). Namely, there are just
$N-r$ independent components amongst $\sigma _0=\int_0^{2\pi l}\sigma dx$
where $N= \dim G,\ r=rank\ G=\dim H$.

Thus, the system has $r$ physical degrees of freedom. However, $PS_{ph}$
does not coincide with ${\bf R}^{2r}$ because there remain discrete gauge
transformations which cannot decrease a number of physical degrees of
freedom, but they do reduce their phase space.

Consider a root system in $H$. Let $P$ be a subset of simple roots
\cite{hel}. A number of simple roots is equal to $r$. $P$ forms a
non-orthogonal basis in $H$ \cite{hel}. All transformations of $h$ being
compositions of reflections $\hat{s}_\omega $ in hyperplanes orthogonal to
simple roots, $(h,\omega )=0$ $\footnote{As $H\sim {\bf R}^r$, one can
assume $(\omega ,h)=\omega _ih_i,\ i$ labels Cartesian coordinates of
the vectors $\omega $ and $h$.}$, $ \omega \in P$, form a subgroup of $G$
called the Weyl group $W$ \cite{hel}, \cite{weyl},
\begin{equation}
\hat{s}_\omega h=\Omega _\omega h\Omega _\omega ^{-1} = h -
\frac{2(h,\omega )}{(\omega ,\omega )}\omega ,\ \Omega _\omega \in G\ .
\end{equation}
Therefore, the points $\hat{s}_\omega p_h,\ \hat{s}_\omega h$ in ${\bf
R}^{2r}$ ( meaning $H\sim {\bf R}^{r}$) should be identified in accordance
with (1) \cite{weyl}. The Weyl group simply transitively acts on the set
of Weyl chambers \cite{hel}, p.458. Any element of $H$ can be obtained
from an element of the Weyl chamber $K^+$ ($h\in K^+$ if $(h,\omega )>0$ for
all $\omega \in P$) by a certain transformation from $W$. The group $W$
does not cover the whole admissible discrete gauge arbitrariness.

Set $E=p_h$ and $A=h$ in (4) and consider gauge transformations with
$\Omega =\Omega _\eta =\exp x\eta/l,\ \eta\in H$. The element $\eta$ cannot
be arbitrary since the periodicity requires $\exp(2\pi \eta)=e,\ e$ the
group unit. This yields
\begin{equation}
\eta=\sum\limits_{\omega \in P}\frac{2n_\omega }{(\omega ,\omega)}\omega
\ ,\ \ \ \ n_\omega \in {\bf Z}
\end{equation}
(a consequence of Lemma 7.6 in \cite{hel}, p.317). Gauge transformations
with $\Omega =\Omega _\eta$ transfer $h$ to $h -a_0\eta$, $a_0=(gl)^{-1}$,
and leave $p_h$ untouched. Semidirect product of a group of these
translations and the Weyl group is called the affine Weyl group $W_A$
\cite{hel}. The discrete gauge arbitrariness is exhausted by $W_A$. Any
element of $W_A$ is a composition of reflections $\hat{s}_{\omega ,n} $ in
hyperplanes $(h,\omega )=a_0n_\omega $\cite{hel}. Thus, the physical
phase space is the quotient
\begin{equation}
PS_{ph}={\bf R}^{2r}/W_A\ ,
\end{equation}
where the action of $W_A$ on ${\bf R}^{2r}$ is determined by all possible
compositions of
\begin{equation}
\hat{s}_{\omega ,n}p_h=\hat{s}_\omega p_h\ ,\ \ \ \ \ \
\hat{s}_{\omega ,n}h=\hat{s}_\omega h + \frac{2n_\omega a_0}{(\omega
,\omega )}\omega \ ,
\end{equation}
where $\hat{s}_\omega \in W$ (cf. (7)), $\omega $ ranges over $P$,
$n_\omega \in{\bf Z}$.

The fundamental modular domain  $K^+_A=H/W_A$ being $CS_{ph}$ in the model
is compact and coincides with the Weyl cell \cite{hel}, a cell of a lattice
vertices of which  are intersection points of hyperplanes $(h,\alpha
)=a_0n,\ n $ an integer, $\alpha $ runs over positive roots. This lattice
is dual to the root lattice with vertices $a_0\sum_Pn_\omega \omega $. For
example, for $SU(3)$ $K^+_A$ is an equal-side triangle with side length
$a_0$. Notice that the size of the fundamental modular domain
non-perturbatively depends on the coupling constants $a_0\sim 1/g$, i.e.,
the geometry of the gauge orbit space $CS_{ph}$ becomes important in a
non-perturbative region \cite{baal}, \cite{zw}.

Let $G=SU(2)$, then $r=1,\ W={\bf Z}_2,\ H\sim {\bf R},\
(\omega ,\omega )=1$. Let us construct $PS_{ph}$.
After identification points $p_h, h+2a_0n,\ n\in {\bf Z}$, on the phase
plane, we get a strip $p_h\in {\bf R},\ h\in [-a_0,a_0]$ with identified
lines $h=\pm a_0$ (a cylinder). On this strip, one should stick together
points $p_h,h$ and $-p_h,-h$. This turns the strip into a half-cylinder
ended by two conic horns (the points $p_h=0, h=0,a_0$).

So, $PS_{ph}$ in 2D Yang-Mills theories is a non-homogeneous
(hypercylinder-like) symplectic space with (hyper)conic singular points
("hyperhorns"). Notice that in the Abelian case $PS_{ph}$ is a cylinder with no
singular points at all. For groups of rank 2, all singular points of
$PS_{ph}$ lie on a triangle being the boundary $\partial K^+_A$ of the
Weyl cell. In a neighbourhood of each point $h_0,p_h=0\in\partial K^+_A$
except the triangle vertices, $PS_{ph}$ locally coincides with $
{\bf R}^2\otimes cone$, where ${\bf R}^2$ is spanned by coordinates
varying along lines $(\alpha ,h-h_0)=a_0\delta _\alpha ,
\ (\alpha ,p_h)=0$, $\alpha $ a root
orthogonal to $\partial K^+_A$, $\delta _\alpha =0$ if $\alpha \in P$ and
$\delta _\alpha =1$ otherwise, while the cone is spanned by coordinates
ranging over lines perpendicular to the above ones. It can be
easily seen from (10) that the second pair of the local canonical coordinates
changes its sign under the reflection $\hat{s}_{\alpha,\delta _\alpha }$
in the straight line containing a part of $\partial K^+_A$ at $h=h_0$,
whereas $\hat{s}_{\alpha ,\delta _\alpha }$ leaves the first  canonical
pair untouched. At the triangle vertices, two conic singularities going
along two edges stick together. If those edges are orthogonal, $PS_{ph}$
looks locally like $cone\otimes cone$, if not, we get a 4D-hypercone as
$PS_{ph}$ in a neighbourhood of the triangle vertices.

A generalization of this singular point pattern in $PS_{ph}$ to groups of
an arbitrary rank is trivial. The Weyl cell is an $rD$-polyhedron.
 $PS_{ph}$ at the polyhedron
vertices has the most singular $2rD$-$hypercone$ structure. On the
polyhedron edges, it is locally viewed as ${\bf R}^2\otimes
2(r-1)D$-$hypercone$. Further, on the polyhedron faces, being polygons, the
local $PS_{ph}$ structure is ${\bf R}^4\otimes 2(r-2)D$-$hypercone$, etc.

A quantization of such symplectic spaces with singular points might give
rise to difficulties \cite{mas}. Fortunately, we know the origin of the
singularities $-$ constraints and gauge symmetry. Therefore, quantization
before reducing phase space (Dirac's operator method) looks preferable.

{\bf 3}. This point is to describe briefly the PI formula
for 4D Yang-Mills theory which takes into account a true geometry of
$PS_{ph}$. Then we shall apply it to the 2D case to verify our general
recipe.

Consider quantum Yang-Mills theory in the Hamiltonian functional
representation, i.e., eigenstates
$\Phi _n$ of the Hamiltonian $H=\langle{\bf
E, E}\rangle/2+\langle{\bf B,B}\rangle/2,\
{\bf E}=-i\delta /\delta {\bf A},\ {\bf B}$ the
colour magnetic field, $\langle , \rangle
=\int d^3x(,)$, are functionals of the vector
potential. The constraint operator (3) (where $A,E\rightarrow {\bf A,E}$)
must annihilate physical states $\sigma \Phi _n=0$ \cite{dir}, which means
that $\Phi _n[{\bf A}^\Omega ]=\Phi _n[{\bf A}]$ (see (4)).

Let $[{\bf A}]$ be a functional space where $\Phi _n$ are defined. Then
$CS_{ph}=[{\bf A}]/G=K$, $G$ acts in $[{\bf A}]$ as (4). Suppose we wish to
parametrize $K$ by fields satisfying a gauge condition $F[{\bf A}]=0,\
{\bf A}\in [{\bf A}]_F\subset [{\bf A}]$. The gauge condition is assumed to
be complete, it fixes all {\em continuous}
 gauge arbitrariness. Due to the Gribov
ambiguity, there is no one-to-one correspondence between $K$ and $[{\bf
A}]_F$. The space $[{\bf A}]_F$ contains gauge-equivalent configurations
${\bf A}^s=\Omega _s{\bf A}\Omega _s^{-1}+g^{-1}\Omega _s\partial\Omega
_s^{-1} $ with ${\bf A},{\bf A}^s\in [{\bf A}]_F$. The set of residual
gauge transformation $S_F=\{\Omega _s\}$ is analogous to $W_A$ in p.2, but
$S_F$ is not always a group \cite{lec} since a composition of two $\Omega
_s$'s does not always give a new copy of ${\bf A} $. Obviously, $K\sim
[{\bf A}]_F/S_F$.

The condition $\sigma \Phi _n=0$ guarantees that $\Phi _n$ are
functionals on $K$. Therefore, one should incorporate somehow the gauge
condition $F=0$ in the Dirac operator method to make sure that a
projection of $\Phi _n$ on the space $[{\bf A}]_F\neq K$ does not break
down the gauge invariance. To reduce the Schroedinger equation $H\Phi _n=
E_n\Phi _n$ on $[{\bf A}]_F$ in a gauge-invariant way, we propose to
introduce curvilinear coordinates \cite{gf}
\begin{equation}
A_j=A_j[a,w]=\Omega \tilde{A}_j\Omega ^{-1}+g^{-1}\Omega
\partial_j\Omega ^{-1}\ ,
\end{equation}
where $\Omega =\Omega [w]$ and $\tilde{A}_j=\tilde{A}_j[a]$ such that
$F[\tilde{A}]\equiv 0$ for all $a\in [a]$, i.e., variables $a$
parametrize the space $[{\bf A}]_F$; by definition $\delta w=\Omega^{-1}
\delta \Omega ,\ \delta $ stands for a functional variation. One
should emphasize that gauge transformations leave $a$ invariant, while
$w$ is transferred by them. Thus, we do not fix a gauge at all, but we do
choose a parametrization of the orbit space by gauge-invariant variables
$a$.

The metric tensor in new variables reads \cite{gf},\cite{lec}
\begin{equation}
\langle \delta {\bf A},\delta {\bf A}\rangle =\langle \delta
q^c,\hat{g}_{cb}[a]\delta q^b \rangle\ , \ \ \ c,b=1,2\ ,
\end{equation}
where $\hat{g}_{cb} $ is a linear operator depending on $F$, $\delta
q^1=\delta a,\ \delta q^2=\delta w$. Rewriting $\delta /\delta {\bf A} $ via
$\delta /\delta q^b$ one can find that $\sigma \sim \delta /\delta w$ and,
therefore, $\Phi _n[{\bf A}]=\Phi_n [\tilde{\bf A}]=\Phi _n^F[a]$, $\Phi
_n^F$ are {\em regular} solutions to the Schroedinger equation
\begin{equation}
\hat{H}_{ph}^F\Phi _n^F=\left( \frac 12 \langle p_a,\hat{g}^{11}p_a
\rangle + V_q[a] +\frac 12 \langle {\bf B},{\bf B} \rangle\right)\Phi _n^F=
E_n\Phi _n^F\ .
\end{equation}
Here $p_a=-i\mu^{-1/2}\delta /\delta a\circ \mu^{1/2},\
\hat{g}_{cb}\hat{g}^{bd}=
\delta _c^d$ and $\mu=(\det \hat{g}_{cb})^{1/2} $; an operator ordering
correction to the potential is
$V_q=1/2\mu^{-1/2}\langle \delta /\delta a, \hat{g}^{11}\delta /\delta
a\mu^{1/2}  \rangle$.

The scalar product is defined in a standard way (the Jacobian $\mu[a]$
has to be taken into account in the scalar product measure \cite{lus})
\begin{equation}
\int_{[{\bf A}]}D{\bf A}|\Phi _n|^2=\int_GDw\int_KDa\ \mu\ |\Phi _n^F|^2
\rightarrow \int_KDa\ \mu\ |\Phi _n^F|^2\ ,
\end{equation}
where an infinite constant $\int_GDw$ can be removed by renormalizing
physical states, which is symbolized by the arrow in (14).
The integration
domain for $a$ has to coincide with $K$. Indeed, consider transformations
of $w$ and $a$ induced by $S_F$, $\Omega [w]\rightarrow \Omega \Omega
^{-1}_s=\Omega [w_s]$ and $\tilde{\bf A}[a]\rightarrow\tilde{\bf A}^s[a]=
\tilde{\bf A}[a_s]$ ( as $\tilde{\bf A}^s\in [{\bf A}]_F$, there exists
$a_s=a_s[a]\equiv \hat{s}a,\ \hat{s}\in S_F$, such that $\tilde{\bf A}^s[a]=
\tilde{\bf A}[a_s]$). Obviously, $A_j[a,w]=A_j[a_s,w_s]$. Hence, the
mapping (11) is one-to-one (i.e., it defines a change of variables) if
$a\in [a]/S_F\sim K$.

Any regular solution to (13) must be automaticly $S_F$-invariant because
$\Phi _n^F[a_s]=\Phi _n[\tilde{\bf A}^s[a]]=\Phi _n[\tilde{\bf A}[a]]
=\Phi _n^F[a]$ where $\Phi _n$ are gauge-invariant regular functionals in
$[{\bf A}]$. So, we do not need to require additionally the
$S_F$-invariance of physical states.

For reasonable gauges $F$ the operator $\hat{g}_{cb}$ is invertible
\cite{zw}. The main goal of our construction is that the scalar product
(14) (amplitudes) and the spectrum $E_n$ do not depend on the choice of
$F$, a change of $F$ corresponds to a change of variables $a\rightarrow
\tilde{a}[a]$. Quantum theories with different $F$'s are unitary
equivalent \cite{lec}.

To obtain the PI representation for the evolution operator kernel
\begin{equation}
U_t^{ph}[a,a']=\langle a|e^{-itH_{ph}^F}|a' \rangle=\sum_n\Phi _n^F[a]
\Phi _n^{F*}[a']e^{-itE_n}\ ,
\end{equation}
one can use the standard slice approximation procedure with the scalar
product (14). A naive implement
of this scheme leads to PI with the integration domain reduced to
$K\subset [a]$. After removing the slice regularization, we get the
problem of treating (or calculating) PI over $K$. For example, in the model
considered in p.2 $K=K^+_A$ is compact. Needless to say, even a finite
dimensional Gaussian integral cannot be explicitly done over a compact
part of an Euclidean space.

The idea to reduce the integration domain in PI for Yang-Mills theory
to the fundamental modular domain has
been proposed in \cite{gr} and developed in \cite{zw}. One can consider
such a recipe as a quantization postulate. However, the analysis of exact
solvable gauge models \cite{uf}-\cite{weyl} has shown that the Dirac operator
formalism leads to another PI representation. Based on this, we propose
the following PI formula \cite{gf}
\begin{eqnarray}
U_t^{ph}[a,a']&=&
\int\limits_{[a]}\frac{Da''}{(\mu\mu'')^{1/2}}U_t^{eff}[a,a'']Q[a'',a']\
,\\
Q[a'',a']&=&\sum_{S_F}\delta [a''-\hat{s}a']\ ,\ \ \ a''\in [a],\ a'\in K\ ,\\
U_t^{eff}[a,a'']&=&\int\limits_{[a]}\prod\limits_{\tau =0}^t
\left( Da(\tau )Dp_a(\tau )\right) \exp i\int\limits_0^td\tau (\langle
p_a,\dot{a} \rangle -H^{eff})\ ,
\end{eqnarray}
where $\mu''=\mu[a'']\ , \int_{[a]}Da'\delta [a-a']\Phi [a']=\Phi [a]$, and
$H^{eff}$ is obtained from the operator $H_{ph}^F$ by replacing the
operator $p_a$ by a c-number and by adding an operator ordering term
$-i\langle \delta /\delta a\ \hat{g}^{11},p_a \rangle/2$. Some details of
deriving (16)-(18) may be found in \cite{lec}.

As follows from (16)-(18), instead of solving the problem of definition
of PI over $K$, one has to calculate PI (18) with the ordinary measure
and then to symmetrize the result with respect to $S_F$-transformations.
The bellow application of (16)-(18) to 2D Yang-Mills theory allows us to
verify precisely our proposal. All functional integrals entering
into (16)-(18) can be explicitly done. The result therefore can be
compared with the known (gauge-invariant) operator
solution of the problem \cite{raj}.
We demonstrate in p.4 that they coincide.

{\bf 4}. For sequent calculations we shall use the Cartan-Weyl basis in $X$
\cite{hel} $[e_\alpha ,e_{-\alpha }]=\alpha $, $\alpha $ a positive root,
$[e_\alpha ,e_\beta ]=N_{\alpha \beta }e_{\alpha +\beta }$ and
\begin{equation}
[h,e_{\pm \alpha }]=\pm (h,\alpha )e_{\pm \alpha }\ ,
\end{equation}
where $h\in H$ and $N_{\alpha \beta }$ are non-zero numbers if $\alpha
+\beta $ is a root. As follows from the analysis of p.2, the conditions
$\partial A=0$ (the Coulomb gauge) and $(e_{\pm \alpha },A)=0$ fix all
continuous
gauge arbitrariness. Therefore one can set $\tilde{A}=h\in H$ in (11),
i.e. $h$ plays the role of gauge-invariant variables spanning $CS_{ph}$.
The mapping (11) determines a change of variables if $h\in K^+_A$ and the
Cartan subalgebra components of the homogeneous part of $\delta w$ (or $w$)
identically vanish, $(h,\delta w_0)\equiv 0$ (we use the notation
introduced in (5)). Thus, $S_F=W_A$ and $K=K_A^+$ in our system.

The metric tensor can be obtained from the equality
\begin{equation}
\delta A=\Omega (dh-g^{-1}\nabla (h)\delta w)\Omega ^{-1}\ .
\end{equation}
Substituting it into (12) we find $\hat{g}_{11}=1,\
\hat{g}_{12}=\hat{g}_{21}=0$ (since $\nabla (h)dh\equiv 0$) and
$\hat{g}_{22}=-g^{-2}\nabla ^2(h)$. The measure $\mu$ in the scalar
product (14) is proportional to $\det\nabla ^{ab}(h)$, where $\nabla
^{ab}(h)=\delta ^{ab}\partial + h^{ab},\ h^{ab}=-gh_if^{iab},\ f^{abc} $
are the structure constants of $X$, indices $i,j$ stand for the Cartan
subalgebra components. Obviously, $h^{ia}=0$. The determinant has to be
calculated on a subspace of ${\cal F}[S^1]$ defined by $f^i_0\equiv 0$
since $\delta w_0^H\equiv 0$. The operator has no zero modes on this
subspace if $h\in K^+_A\ , (\partial K^+_A$ is not included into $K^+_A$).

The operator $\nabla (h)$ acts as an infinite dimensional matrix in the
space of Fourier coefficients $f_0^{\pm \alpha },\ f^i_{c,sn},\ f^{\pm
\alpha }_{c,sn} $ where the upper index $\pm \alpha $ symbolizes
components corresponding to the basis elements $e_{\pm \alpha } $. This
infinite matrix has the block-diagonal form, each block is finite
dimensional. We denote these blocks $\nabla_0,\ \nabla _n^H,\
\tilde{\nabla }_n \ (n=1,2,...)$. They act in invariant subspaces of
$\nabla(h)$ composed of coefficients $\{f_0^{\pm \alpha }\},\
\{f_{s,cn}^i\},\ n $ fixed, $\{f_{c,sn}^{\pm \alpha }\},\ n$ fixed,
respectively. Hence, $\det\nabla =\det\nabla_0\prod_1^\infty(\det
\nabla_n^H \det\tilde{\nabla}_n) $. Straightforward calculations in the
Cartan-Weyl basis (actually, only (19) is sufficient to use) lead
to the following result $\footnote{The Cartan-Weyl basis is not orthogonal.
Its connection with a real orthogonal basis in $X$ is given in \cite{jac},
p.149. The latter is important for a correct calculation of $h^{ab}$.}$,
$\det\nabla = C(l)\mu(h),\ \mu=\kappa^2(h)$ and
\cite{gra}, p.37,
\begin{equation}
\kappa(h) =\prod\limits_{\alpha >0}\left[\frac{\pi (h,\alpha )}{a_0}
\prod\limits_{n=1}^\infty\left(1-\frac{(h,\alpha )^2}{a_0^2n^2}
\right)\right] =\prod\limits_{\alpha >0}\sin\frac{\pi (h,\alpha )}{a_0}\ .
\end{equation}
The infinite constant $C(l)$ appears to be included into a definition of
the symbol $Dw$, i.e., into the norm of physical states.

Notice also that $\det\nabla (h)$ coincides with the Faddeev-Popov
determinant in the above (Coulomb) gauge condition (cf. the $SU(N)$ case
considered in \cite{het}). It is positive on the fundamental modular domain
$K^+_A$ and vanishes at its boundary in accordance with a general
analysis \cite{zw}. Moreover, the vacuum configuration $A=h=0$ coincides
with the most singular points of $\det\nabla$, it vanishes as
$|h|^{2N_+}$, $N_+=(N-r)/2$ a number of positive roots.

The physical Hamiltonian reads
\begin{equation}
H_{ph}=-\frac{\hbar^2}{4\pi
l}\frac{1}{\kappa(h)}(\partial_h,\partial_h)\circ\kappa(h) -E_0\ ,
\end{equation}
where we restore the Planck constant and take into account
$\hat{g}^{11}=1$ in (13); the factor $(2\pi l)^{-1}$ results from $\delta
/\delta A=(2\pi l)^{-1}\Omega \partial_h\Omega ^{-1}+...$ and integration
over $x$;
\begin{equation}
V_q = \frac{\hbar^2}{4\pi l}\kappa^{-1}\partial_h^2\kappa
=-\frac{\pi\hbar^2}{4a_0^2l}\left(\sum\limits_{\alpha >0}\alpha
\right)^2\equiv -E_0\ .
\end{equation}
The proof of (23) is analogous to that given in \cite{weyl} to show
vanishing the operator ordering corrections in the model considered there.

The effective Hamiltonian in (18) for the operator (22) coincides with a
Hamiltonian of an $r$-dimensional free particle with mass $2\pi l$. Doing
the PI (18) we derive from (16) our final result
\begin{equation}
U_t^{ph}(h,h')=\left(\frac{l}{i\hbar t}\right)^{r/2}
\sum\limits_{\hat{s}\in W_A}(\kappa (h)\kappa (\hat{s}h'))^{-1}
\exp\left(\frac{i\pi l(h-\hat{s}h')^2}{\hbar t} +itE_0\right)\ ,
\end{equation}
where we have put in (16) $[a]={\bf R}^r,\ \mu^{1/2}=\kappa ,\ S_F=W_A$
and done the integral over $a''\equiv h''$.

One should stress that, first, the PI thus constructed obeys the
convolution rule (see (14))
\begin{equation}
U_{t+t'}^{ph}(h,h')=\int_{K^+_A}dh''\kappa
^2(h'')U_t^{ph}(h,h'')U_{t'}^{ph}(h'',h')
\end{equation}
and, second, it gives a regular solution to Schroedinger equation
$(i\hbar\partial_t-H_{ph})U_t^{ph}=0$. The latter means,
as has been argued in p.3, that the amplitude (24) must be an analytical
gauge-invariant functional in the whole configuration space $[A]$. We
shall see bellow that this is so.

As an independent test, one can also recover (24) by summing the spectral
series (15). The {\em regular} eigenfunctions of (22) are $\footnote{For
$G=SU(2)\ 2\pi$ in the exponential has to be replaced by $\pi$ because
the root matrix is a number, $(\omega ,\omega )=1$.}$
\begin{equation}
\Phi _{(n)}(h)=\frac{const}{\kappa (h)}\sum_{\hat{s}\in W}\det\hat{s}
\cdot \exp\frac{2\pi i}{a_0}(\gamma _{(n)},\hat{s}h)\ ,
\end{equation}
where the vector $\gamma _{(n)}=\sum_Pn_\omega \omega ,$ belongs to the
root lattice ($n_\omega $ integers), by definition $\det\hat{s}_\omega
=-1$ and $\det\hat{s}\hat{s}'=\det\hat{s}\det\hat{s}'$. The sum of
exponents in (26) has zeros (if it does not identically vanish at a given
$\gamma _{(n)}$) of the same order as $\kappa $ has on $\partial K_A^+$.
The regular solutions (26) turn out {\em automaticly} to be invariant
under transformations from the affine Weyl group. The invariance under
translations $h\rightarrow h+a_0\eta$ is guaranteed by that $-2(\omega
,\omega ')/(\omega ,\omega )$ is an integer for $\omega ,\omega '\in P$
\cite{hel}. The W-invariance can be seen from $\kappa (\hat{s}_{\omega
,n}h)=-\kappa (h),\ \omega \in P,$ and $(\hat{s}\gamma _{(n)},\hat{s}h)=
(\gamma _{(n)},h)$ for any $\hat{s}\in W$. Thus, {\em no} additional
requirement of the $W_A$-invariance must be imposed on physical states
(cf. p.3).

Not all of the functions (26) are linearly independent at fixed $\gamma
_{(n)}^2$ (energy level). Due to the W-invariance, the transformations
$\gamma_{(n)}\rightarrow \hat{s}\gamma _{(n)},\ \hat{s}\in W,$ leave (26)
unchanged. If $\gamma _{(n)}\in \partial K^+$, then $\Phi _{(n)}\equiv 0$.
Indeed, the sum (26) changes its sign under any reflection from $W$. On
the other hand, consider a reflection $\hat{s}_0$ in a hyperplane of
$\partial K^+$ where $\gamma _{(n)}$ lies. Obviously, $\hat{s}_0\gamma
_{(n)}=\gamma _{(n)}$ and, hence, $\hat{s}_0$ leaves the sum untouched,
which is possible only if the sum is zero. So, the spectrum reads
\begin{equation}
E_{(n)}=\frac{\pi \hbar^2}{a_0^2l}\gamma _{(n)}^2 -E_0,\ \ \
\gamma_{(n)}=\sum_Pn_\omega \omega \in K^+\ .
\end{equation}
The root lattice vertices belonging to the Weyl chamber $K^+$ determine
the irreducible representations of $G$ \cite{jac}. The operator (22) is,
in fact, proportional to the quadratic Casimir operator of $G$. Hence,
the eigenvalues (27) should be also proportional to values of the Casimir
on the characters of the irreducible representations of $G$ $\footnote{If
for instance $G=SU(2)$, one can always set $(\omega ,\omega )=1$ and
$\gamma _n=\omega n,\ n$ a positive integer ($K^+ $ is a positive
semiaxis), then the (Casimir) spectrum (and the irreducible
representations) is labelled by the spin $j=0,1/2,1,...,\
E_n=E_0(n^2-1)=4E_0j(j+1),\ E_0=\pi \hbar^2/(4a_0^2l)$.}$. The
eigenfunctions are therefore expressed via these characters, i.e., via
$Tr\exp (2\pi h/a_0)=TrW[A]$ \cite{mig},\cite{raj}. The latter results in
the explicit gauge invariance of the transition amplitude (24) because it
can be expanded into the spectral series (15).

Thus, the PI constructed above recover all results of the loop
(gauge-fixing free) approach for 2D Yang-Mills theory
\cite{mig},\cite{raj}. Lessons following from our consideration are that,
first, one should {\em not} reduce the integration region in PI to the
fundamental modular domain (the Weyl cell in the model). The correct PI is
obtained by summing over all trajectories reflected from the boundary
$\partial K^+_A$ (Gribov's horizon) and connecting the initial and final
configurations. It resembles the PI quantization of a particle on a circle
(or in a box). The reduction of the integration domain in PI to an
interval is known to be meaningless \cite{pau}. Secondly, in Dirac's
operator approach, no additional $W_A$-invariance condition
must be imposed on physical states (compare with \cite{sem}, \cite{het}).
Thirdly, the spectrum strongly depends on the $PS_{ph}$ geometry. If one
assumes $PS_{ph}$ to be ${\bf R}^{2r}$, the spectrum would be continuous,
which is wrong.

{\bf 5}. Including fermions in our PI approach does not meet serious
difficulties. One should add the standard gauged fermion Hamiltonian to
(2) and modify the Gauss law (3) $\sigma \rightarrow \sigma +\rho (\psi
,\psi ^+)$ with $\rho $ being the colour charge density of the fermion
field $\psi $ \cite{mic},\cite{sem}. To solve the constraint $\sigma \Phi
_{ph}=0$, one has to introduce curvilinear coordinates on the superspace
$[A,\psi ]$ (the Grassmann holomorphic representation for fermion
operators is assumed to be used) \cite{jp}. Their bosonic part coincides
with (11), while the fermionic one is $\psi =\Omega \xi $ where the
fermion field $\xi $ plays the role of physical fermion variables. Then,
$\sigma \Phi _{ph}=0$ decouples into two independent parts $\delta
/\delta w \Phi_{ph}=0$ and $\rho_0^H(\xi ,\xi ^+)\Phi _{ph}=0$ where
$\rho _0^H$ is a homogeneous component of $\rho $ (cf. (5)) belonging to
the Cartan subalgebra. We remind that gauge transformations from the
Cartan subgroup $G_H$ of $G$ leave $h$ untouched, but they do transform the
fermion field $\xi $. The $G_H$-invariance of physical states yields
the constraint $\rho _0^H\Phi_{ph}=0$.

The Laplace-Beltrami operator $\langle \delta /\delta A,\delta /\delta A
\rangle$ in curvilinear supercoordinates contains the term $V_f=\langle
\tilde{\rho },\nabla^{-2}(h) \tilde{\rho } \rangle/2$ in addition to (22),
$\tilde{\rho }=\rho -\rho _0^H$ (the terms with $\delta /\delta w$ and
$\rho _0^H$ vanish on physical states). The operator $\nabla^{-2}(h)$ can
be obtained in the same fashion as we have calculated $\det\nabla(h)$.
Regular solutions to the functional Schroedinger equation on the physical
subspace will automaticly be $W_A$-invariant \cite{jp}. Notice that the
group $W_A$ non-trivially acts on fermion fields and identifies some points
on the Grassmann hyperplane $\xi ,\xi ^*$ ($PS_{ph}$ of physical fermions
degrees of freedom is modified \cite{uf},\cite{jp}). The PI
formula has the same form $\hat{U}_t^{ph}=\hat{U}_t^{eff}\hat{Q}$
\cite{gf},\cite{lec}, i.e., it contains the $W_A$-symmetrization provided
by $\hat{Q}$ rather than the integration domain reduction. The PI for
$\hat{U}_t^{eff}$ cannot be explicitly done because of the presence of
the non-Gaussian term $V_f$ in the effective action.

An important feature appeared in the mixed model is that the operator
$\hat{Q}$ simultaneously symmetrize both gauge and fermion fields, $h$
and $\xi $, with respect to $W_A$. Therefore $\hat{Q}$ does not commute
with fermion creation and destruction operators, which might result in a
modification of the fermion Green functions in a non-perturbative region
\cite{gf},\cite{lec}. Also, the residual transformations responsible for
translations $h\rightarrow h+a_0\eta$ mix fermion creation and
destruction operators. It leads to the anomalies in the model \cite{sem}.

A detail derivation of the modified PI representation for the 2D
Dirac-Yang-Mills theory will be given elsewhere together with an
investigation of dynamical consequences emerging due to the non-trivial
geometric structure of $PS_{ph}$.
\begin{center}

{\bf \Large Acknowledgments}
\end{center}

The author is kindly grateful to Prof. J.Klauder (Florida University) and
Prof. A.Di Giacomo (Pisa University) for the warm hospitality extended to
him during his stay in Gainesville and Pisa where this work has been done.

\end{document}